# Structure, Risk, and Access to Credit: Reassessment of the Paycheck Protection Program Effectiveness

Chunyu Qu[*]

Dec 17, 2025

**Abstract**

The Paycheck Protection Program (PPP) was the largest targeted business support program in U.S., yet its firm-level effects remain contested. I link administrative PPP and SBA 7(a) records to a near-universe panel of U.S. employer firms from Dun & Bradstreet, covering roughly 30 million establishments, and evaluate short-run impacts along employment, financial stress, and commercial credit risk. To address non-random take-up, I combine PSM with difference-in-differences on a balanced panel from March to September 2020 and exploit variation in loan holding duration.

PPP receipt raises employment by about 0.07 percent on average but improves failure-risk and delinquency-risk percentile rankings by roughly 1.2 and 3.2 points, respectively, with longer loan duration strengthening all three margins. Heterogeneity analysis shows that small–medium firms and borrowers with intermediate pre-crisis risk experience the largest gains, while micro, very large, and highly stressed firms benefit less. Firms without prior 7(a) borrowing relationships realize particularly large credit-score gains.

Overall, the evidence indicates that PPP functioned more as a balance-sheet and credit-risk backstop than as a powerful jobs program for the average treated firm. The results highlight how firm structure, pre-crisis financial health, and access to government-backed credit shape the effectiveness of large-scale emergency support.

---

[*] Chunyu Qu holds a Ph.D. in Economics from Fordham University and is currently a Data Scientist with Dun & Bradstreet Inc (D&B), 2900 Esperanza Crossing, Austin, TX 78758 (email; cqu9@fordham.edu).
The author thanks Sophie Mitra for her major feedback throughout the research. He also thanks Philip Shaw, Subha Mani, Andrew Simons, Patricia Gomez-Gonzalez, and Johanna L. Francis from Fordham University for their constructive feedback. The author further acknowledges Amber Jaycocks, Nalanda Matia, Nivedh Nijayakrishnan, Julia Zhao, Michelle Solomon, and Andrew Bynes at Dun & Bradstreet for their support, and Tuhin Batra, formerly of D&B and now with Atlantic Technological University, Ireland, for their technical insights and helpful discussions. The author thanks Qi He of Google LLC for her insights of the policy effectiveness and financial stability metrics. Her expertise in enterprise-level financial planning and cost-efficiency analytics informed the interpretation of firm-level credit and liquidity responses under large-scale policy interventions. *All views expressed in this paper are solely of the author and do not represent the views of D&B or any affiliated institution.*

# Introduction

The Paycheck Protection Program (PPP) was the largest targeted business support program in U.S. history. Between April 2020 and May 2021 it disbursed more than $800 billion in forgivable loans to small and medium-sized firms to prevent layoffs and business failures during the COVID-19 shock (Hubbard and Strain, 2020; Autor et al., 2021a). A rapidly growing literature evaluates whether PPP preserved employment, supported firm survival, and stabilized local economies (Bartik et al., 2020; Faulkender et al., 2020; Autor et al., 2021a, 2021b; Dalton, 2021; Doniger and Kay, 2021; Barraza et al., 2020; Cororaton and Rosen, 2021; Wang et al., 2021). These studies generally find that PPP cushioned job losses and business distress, but at high fiscal cost, with effects that faded over time and with imperfect targeting across firms, lenders, and communities (Granja et al., 2020; Humphries et al., 2020; Joaquim and Netto, 2021; Griffin et al., 2021; Atkins et al., 2021; Fairlie and Fossen, 2021; Chernenko and Scharfstein, 2021).

Much less is known about how pre-existing financial health and credit access shaped PPP's effectiveness at the firm level, and how employment effects compare with changes in firms' commercial credit standing. Existing work has largely focused on jobs, survival, and allocation patterns rather than on standardized risk measures used by lenders. This leaves open questions: Did PPP mainly operate through payroll, or did it also shift firms' forward-looking risk profiles? Were gains concentrated among already strong firms or among vulnerable but viable ones? And how did outcomes differ between businesses with established government-backed credit relationships and those previously outside the SBA system?

In this paper, I address these questions using a large-scale linkage between PPP public records and Dun & Bradstreet's (D&B) commercial data. I construct an establishment-level panel that covers essentially the universe of active U.S. employer firms and match PPP loans to D&B information on employment, industry, location, and two forward-looking risk scores: the Financial Stress Score (FSS), which summarizes the predicted probability of business failure, and the Commercial Credit Score (CCS), which captures trade-credit and payment performance. My outcome variables are log employment and the percentile ranks of FSS and CCS, allowing me to evaluate not only whether PPP stabilized payrolls, but also whether it improved firms' financial health and creditworthiness.

To address non-random program take-up, I combine PSM with a DID specification. First, I estimate the propensity to receive a first-draw PPP loan using rich pre-treatment covariates, including firm size, age, industry, geography, pre-PPP risk scores, and prior SBA 7(a) borrowing history, and match PPP recipients to observationally similar non-recipients within strata of the propensity score. Second, on the matched sample I estimate two-period DID models with establishment fixed effects, comparing changes in employment and risk scores between March and September 2020 for PPP and non-PPP firms. I further exploit variation in loan-holding duration to assess how the length of PPP support relates to subsequent outcomes.

The central theme of the paper is how PPP's effectiveness varies with firms' pre-PPP financial conditions and credit access. I systematically stratify effects along three dimensions. First, I examine heterogeneity by firm size and by pre-PPP FSS and CCS categories, capturing differences in scale and underlying financial stress and credit risk. Second, I contrast firms with and without prior SBA 7(a) borrowing relationships, which proxy for access to government-backed credit and established lender ties. Third, I document industry heterogeneity across major NAICS sectors, highlighting how PPP operated in manufacturing, retail and transportation, hospitality, finance and real estate, and public administration.

The main findings are as follows. At the aggregate level, PPP loans generate modest but positive gains in employment, while improvements in financial stress and commercial credit scores are larger in magnitude. These averages mask substantial heterogeneity: firms in the middle of the size distribution and those with intermediate pre-PPP financial stress and credit risk experience the largest improvements in both employment and risk scores, whereas the smallest and largest firms and the least- and most-stressed firms show smaller or statistically weaker responses. This pattern suggests that PPP support was most effective for financially vulnerable but viable firms, for whom liquidity relief translated into both job retention and measurable reductions in predicted failure and delinquency.

I also uncover important heterogeneity related to credit access. Firms with prior SBA 7(a) borrowing relationships benefit from PPP in terms of employment and risk scores, consistent with relationship lenders facilitating access to the program (Joaquim and Netto, 2021). However, non-7(a) firms, those without documented prior SBA loans, display even larger gains in commercial credit scores, indicating that PPP played a critical role in easing trade-credit constraints for previously less-connected businesses. Across industries, PPP effects on credit risk are strongest in retail and transportation, food and entertainment, and public administration, sectors that faced the most severe revenue disruptions, whereas employment effects are more uniform but generally smaller (Bartik et al., 2020; Granja et al., 2020). Loan duration is positively associated with improvements in financial stress and commercial credit, reinforcing the interpretation of PPP as a liquidity hedge rather than a pure payroll subsidy.

This paper makes three main contributions to the PPP literature. First, I provide a unified firm-level evaluation of PPP's short-run effects on employment, financial stress, and commercial credit risk using a panel that closely approximates the universe of U.S. employer firms and leverages standardized D&B risk scores. Second, I systematically analyze heterogeneity in PPP effects across firm size, pre-PPP financial stress, pre-PPP commercial credit quality, prior SBA 7(a) borrowing relationships, ownership type, and industry, rather than focusing on a single dimension or outcome. Third, by emphasizing how pre-existing financial health and credit access condition the translation of PPP support into both employment and credit improvements, my results complement existing evidence on targeting and racial disparities and yield concrete lessons for the design of future emergency business support programs.

## Section II. Related Literature

A large and rapidly growing literature evaluates the Paycheck Protection Program (PPP) along several dimensions. A first group of studies estimates employment and survival effects using payroll, tax, or survey data (Bartik et al., 2020; Faulkender et al., 2020; Autor et al., 2021b; Dalton, 2021; Doniger and Kay, 2021; Barraza et al., 2020; Cororaton and Rosen, 2021; Wang et al., 2021). These papers generally conclude that PPP preserved a meaningful number of jobs and reduced business failures in the short run, but at high fiscal cost. Autor et al. (2021a, 2021b) and Chetty et al. (through related CARES Act work) estimate that PPP preserved roughly 1–3 million jobs over horizons of 5–14 months, with implied costs per job-year in the $170,000–$377,000 range, and that the effects on employment attenuated as the crisis evolved (Hubbard and Strain, 2020; Autor et al., 2021a). Joaquim and Netto (2021), Wang et al. (2021), and Cororaton and Rosen (2021) link PPP records to Dun & Bradstreet (D&B) or similar files to track business closures, delinquency, and risk scores. They generally find that PPP reduced failure risk and improved certain credit indicators, especially for firms that were financially constrained ex ante.

A second group of studies focuses on allocation, targeting, and efficiency. Granja et al. (2020), Barrios et al. (2020), and Bartik et al. (2020b) show that PPP funds were not tightly targeted to the firms or regions most adversely affected by the pandemic. Larger and better-banked firms, and those in areas with greater bank branch density, were more likely to obtain loans and did so earlier, reflecting information frictions and capacity constraints among lenders (Humphries et al., 2020; Joaquim and Netto, 2021). Several papers document that many firms used PPP funds to service fixed non-payroll costs or to build up cash buffers, leading to smaller-than-expected employment responses but stronger balance-sheet protection (Granja et al., 2020; Cororaton and Rosen, 2021; Doniger and Kay, 2021). This line of work emphasizes that PPP was designed and implemented under severe time pressure, which limited the ability to finely target support and raised questions about cost-effectiveness.

A third strand examines distributional and racial disparities in PPP access and outcomes. Using SBA and bank-level data, Atkins et al. (2021), Chernenko and Scharfstein (2021), and Fairlie and Fossen (2021) find that Black-owned and other minority-owned businesses received smaller loans, were less likely to obtain PPP support in the early rounds, and relied disproportionately on fintech and non-bank lenders. Journalistic and policy reports reach similar conclusions about delayed access and smaller loans for minority-owned firms (Beer, 2020; Cowley, 2021; Rosenberg and Myers, 2020; Zhou, 2020). Fintech and small-bank lenders partially mitigated these gaps by serving minority-owned firms that were less likely to be approved by traditional banks (Erel and Liebersohn, 2020; Griffin et al., 2021; Cross, 2021; Crutsinger, 2021; Balyuk et al., 2021). These findings connect to a broader literature on information frictions, relationship lending, and unequal access to credit for small businesses and minority entrepreneurs (Balyuk et al., 2020; Bates and Robb, 2013; Bates et al., 2018).

My study contributes to this literature in three ways. First, I provide a unified firm-level evaluation of PPP's short-run effects on employment, financial stress, and commercial credit risk using a large panel that closely approximates the universe of U.S. employer firms and leverages standardized D&B risk scores. Second, I systematically analyze heterogeneity in PPP effects across firm size, pre-PPP financial stress, pre-PPP commercial credit quality, prior SBA 7(a) borrowing relationships, ownership type, and industry, rather than focusing on a single dimension or outcome. Third, by emphasizing how pre-existing financial health and credit access condition the translation of PPP support into both employment and credit outcomes, I complement existing work on targeting and racial disparities and provide evidence directly relevant for the design of future emergency business support programs.

# Section III. Data

In this section, I describe the data sources, construction of key variables, and summary statistics used in the analysis. The empirical work combines Dun & Bradstreet's commercial database with loan-level information from the Small Business Administration (SBA) on PPP and 7(a) lending.

## A. Data Sources

### 1. Dun & Bradstreet commercial data

Dun & Bradstreet (D&B) maintains the largest commercial database of private businesses in the United States. Each establishment is identified by a unique DUNS number and is tracked monthly on employment, industry, location, sales, ownership, and several forward-looking risk scores. These scores are widely used by lenders, suppliers, and public agencies for credit risk assessment and portfolio management.

For this paper, I use D&B's monthly panel for U.S. establishments on the period from January 2019 through June 2021. In June 2021, D&B records approximately 33.9 million businesses, of which about 33.8 million have no more than 500 employees, and roughly 21.0 million are classified as "small businesses" based on D&B's calibration to SBA standards. Over the same period, SBA reports about 31.7 million small businesses, of which 27 million are non-employers. The difference in counts reflects that many sole proprietors and micro businesses do not maintain a DUNS profile unless they seek external credit. Since PPP loans require a DUNS number or a relationship with a DUNS-registered entity at the time of application, my dataset covers essentially all firms with measurable credit histories.

### 2. SBA PPP data

Loan-level PPP data are obtained directly from the SBA as of August 15, 2021. The dataset includes the business name and address, business type, loan amount, self-reported number of jobs retained, loan approval date, lender (originator), and servicer for every approved PPP loan. I focus on first-draw PPP loans originated during 2020.

I merge the PPP data to the D&B panel using DUNS numbers where available and supplement this with in-house name and address matching based on business name, street address, city, and state. The PPP dataset contains 11,768,689 loan records. In 2020, 5,138,180 loans were approved for 4,815,433 unique DUNS numbers, implying that about 192,000 businesses received more than one first-draw loan under the same DUNS identifier. Multiple loans can arise from legitimate within-firm applications (e.g., separate establishments) or from misreporting and fraud. To avoid double-counting treatment, I collapse duplicate DUNS numbers within 2020 and retain one record per establishment. Time-series patterns of PPP loan approvals by count and dollar volume are documented in Appendix B.

### 3. SBA 7(a) data

The SBA 7(a) program provides government-guaranteed loans to small businesses through banks and non-bank lenders. I use the SBA's public 7(a) dataset, which contains all approved 7(a) and 504 loan guarantees since January 1, 1990 and is updated quarterly. The dataset includes borrower name and address, loan amount, approval date, and lender information, and has roughly 1.5 million entries.

I merge the 7(a) data to the D&B panel using name and address matching, and construct an indicator for whether a firm had any 7(a) loan approved between 2010 and 2019. Restricting to pre-2020 originations avoids simultaneity with PPP and ensures that the 7(a) variable captures established lending relationships and prior access to government-backed credit. This 7(a) history indicator is used as a stratification variable and, in some specifications, as a proxy for pre-existing credit access.

### B. Business Performance and Heterogeneity Variables

All outcome, control, and stratification variables are constructed from the merged D&B–SBA dataset.

#### 1. Business performance (dependent variables)

**Employment**. PPP eligibility is generally based on having no more than 500 employees, but the official rules allow for industry-specific exceptions; detailed eligibility criteria are summarized in Appendix A.

**Financial Stress Score** (FSS). It summarizes the probability that a business will seek legal relief from creditors or cease operations without paying all creditors in full over the next 12 months. The score is provided both as a five-class risk category (1 = lowest risk, 5 = highest risk) and as a 1–100 percentile ranking of failure risk within the national firm population. I use the percentile ranking, where higher values indicate lower predicted failure risk. Appendix Table B.1, Panel A, reports the mapping from raw scores to risk classes and percentiles.

**Delinquency Predictor Score** (Commercial Credit Score, CCS). It predicts the likelihood that a business will pay at least 10 percent of its obligations 91 days or more past due, obtain legal relief, or cease operations without paying all creditors in full within the next 12 months. The scale is the same as the Financial Stress Score. I use the percentile ranking of the DPS as my main credit-risk outcome; Appendix Table B.1, Panel B, summarizes the scale.

#### 2. Control variables - Industry and Sales

Industry is measured using North American Industry Classification System (NAICS2) codes. Annual sales (revenues) in U.S. dollars are reported in the D&B data and updated periodically. I use the logarithm of annual sales in the pre-PPP period as a control for firm scale and revenue magnitude. Within-year monthly observations inherit the same annual sales value.

#### 3. Stratification variables

**Owner race and ethnicity.** I construct an indicator for "ethnic minority ownership," equal to one if the owner is Hispanic, Black or African American, American Indian, Asian, Native Hawaiian, Pacific Islander, Alaska Native, or reports multiple minority races, and zero otherwise. I focus on the binary minority vs. non-minority distinction in the heterogeneity analysis.

**Public vs. private status.** Although PPP was designed primarily for small and closely held businesses, publicly listed firms were also eligible under the statutory rules. I use the public vs. private indicator as a stratification variable when comparing PPP effects across ownership types.

**7(a) financing history.** Using the merged SBA 7(a) dataset, I define a pre-PPP 7(a) history dummy that equals one if a firm received at least one 7(a) or 504 loan between 2010 and 2019, and zero otherwise. This variable proxies for established relationships with SBA-participating lenders and prior access to government-guaranteed credit. It plays a central role in my heterogeneity analysis, where I compare PPP effects between firms with and without prior 7(a) borrowing.

## C. Summary statistics

Summary statistics for the main variables are reported in Table 1. The table shows that PPP recipients are, on average, larger, less risky, and more likely to have prior 7(a) borrowing and minority owners than non-recipients before matching. After PSM, observable differences in employment, sales, risk scores, industry, and location are substantially reduced, providing a balanced sample for the DID analysis described in the next section.

# Section IV. Models

I estimate the impact of PPP loans on employment, financial stress, and commercial credit risk using Propensity Score Matching (PSM) combined with difference-in-differences (DID). This PSM–DID framework mitigates selection into treatment based on observables and controls for time-invariant unobserved firm characteristics. I also implement an event-study specification to trace the dynamic effects of PPP over time.

## A. Identification Strategy

Receiving a PPP loan required not only formal eligibility but also timely information, the capacity to navigate application rules, and relationships with lenders. Information frictions, cash-flow management, financing capabilities, and owner competencies are therefore likely to affect both PPP take-up and subsequent outcomes. These factors are not observed in the data and could confound comparisons between recipients and non-recipients.

A second challenge is that I only observe approved loans. Among non-recipient firms, I cannot distinguish those that applied and were rejected, those that were ineligible, and those that were eligible but did not apply. To address these issues, I adopt a two-step PSM–DID design (Imbens and Wooldridge, 2009). First, I estimate a propensity score using a logistic regression where the dependent variable is an indicator for receiving a first-draw PPP loan. The covariates are baseline characteristics measured in March 2020: industry (NAICS), state, public listing status, ownership type, prior 7(a) loan history, annual sales, and a race dummy for minority-owned businesses. These variables plausibly affect both PPP take-up and subsequent outcomes. Appendix Table B.3 shows that correlations among these covariates are modest ($\leq 0.25$), supporting their use in the propensity model.

Second, I use nearest-neighbor matching on the estimated propensity scores to construct a control group of non-recipient firms with similar pre-PPP characteristics as recipients. I assess balance by comparing the distribution of pre-PPP covariates between treated and matched control firms. All subsequent models are estimated on this matched sample, where "treated" firms are PPP recipients and "control" firms are their matched non-recipient counterparts.

The number of businesses increases over time, so I identify firms by their DUNS number and construct a balanced panel of establishments observed both before and after PPP. In the main PSM–DID specification, I define

March 2020 as the pre-PPP period (t = 0) and September 2020 as the post-PPP period (t = 1). This window spans the first PPP round and the immediate adjustment period, and allows me to assess changes in employment, financial stress, and commercial credit over a complete early PPP cycle rather than only at very short horizons. I first conducted a DID model on the matched controls and treated firms as follows:

$$y_{it} = \gamma PPP_i + \lambda Post + \beta PPP_i \times Post + \rho X_i + \delta_{st} + \delta_{nt} + \epsilon_{it} \quad (1)$$

where $y_{it}$ is the variable measuring the performance of business i at time t, including log of employment, financial stress percentile ranking, and commercial credit percentile ranking. $PPP_i$ =1 if the small business received a loan and 0 if not. Post=0 if the firm data was observed in pre-PPP period, and Post=1 when the firm data was in the post-PPP period. The DID term $PPP_i \times Post$ =1 if an observation on firm i received a loan in the post period. The coefficient β of the interaction term represents the average treatment effect (ATE) on the treated (ATT). $X_i$ is a vector with three controls at $t = 0$ including the log of annual sales (in USD), a binary indicator if the firm imports materials, and a binary indicator if the firm exports goods or services. $\delta_{st}$ is a set of state-by-time fixed effects. This effect captures unobservable factors that vary across states, including regional economic conditions and state-level policies in for different periods. $\delta_{nt}$ is the NAISC 2-digit category industry-by-time fixed effects that capture the industry-specific trends or shocks that change in different times. By including these control variables and fixed effects, my model attempts to isolate the impact of the treatment variable of interest while minimizing potential bias stemming from omitted variable concerns.

Furthermore, I stratify results of model (1) using several variables that can influence both the likelihood of receiving loans and the impact of the loans, including minority ownership dummy; pre-PPP employment levels (logarithm of the employee count, a continuous variable ranging from 1 to 7.31); financial stress (a categorical variable from 1 to 5); commercial credit risk (a categorical variable from 1 to 5); public dummy (coded as 1 if a business sells shares to the public at large in the equity market, and 0 otherwise); NAICS-1 digit industry classification; and 7(a) history dummy (coded as 1 if a business has a 7(a) loan history prior to 2021).To evaluate the treatment-on-the-treated effects of the duration of receiving loans, I conducted a DID model as follows

$$y_{it} = \gamma PPP_i + \lambda Post + \beta Duration_{it} + \rho Xi + \delta_{st} + \delta_{nt} + \epsilon_{it} \quad (2)$$

where $Duration_{it}$ is the duration of PPP benefit receipt in numbers of months from April to August 2020. For example, it is equal to five if a firm receives the loan from April to August, and 1 if a firm receives the loan only in August. Given the time taken by the loan application process, $Duration_{it}$ reflects the timeliness of getting loans. For non-recipients, $Duration_{it}$ is always equal to 0. Besides, all the other terms are the same as in model (1) above.

## B. Parallel Trends

The parallel trends assumption is critical for the identification of the impact of the PPP loans in the DID model. PPP recipients and non-recipients may not follow the same trend in the Pre-PPP period, which might affect DID estimates. I run an event study and a test on the specification as follows:

$$y_{it} = \gamma PPP_i + Month_t + \sum_{t=-4}^{5} \beta_t PPP_i \times Month_t + \rho X_i + \delta_{st} + \delta_{nt} + \epsilon_{it} \quad (3)$$

In the above event study, the balanced panel spans ten months, from November 2019 through August 2020. The pre-PPP months include November 2019 to March 2020, while the post-PPP months consist of April, May, June, July, and August 2020. t is the parameter for corresponding month (dummy variables). The range $t = -4$ to $t = 0$ represents the pre-PPP months, while $t = 1$ to $t = 5$ corresponds to the post-PPP months. The reference category is March 2020 ($t = 0$). All the other terms are the same as specification (1).

To check the parallel pre-trend assumption, I tested the hypothesis: $H_0: \beta_{-4} = 0, \beta_{-3} = 0, \beta_{-2} = 0, \beta_{-1} = 0$, which shows insignificant F-stat. Figure 2 shows the dynamic DID estimators of Equation (3), from which I see there is no significant different between the pre-PPP coefficients for log(employment), commercial credit ranking, and financial stress ranking, indicating no pre-existing trends. Both graphic and statistical evidence suggest that the treatment and control groups exhibited parallel trends in the Pre-PPP period. Hence, any post-PPP divergence in trends between the treatment and control groups may be attributed to the impact of the event under study.

## Section V. Results

This section presents a comprehensive analysis of the overall and stratified impacts of loans on various aspects of business performance.

### A. Descriptive Statistics

Tables 1 and 2 report summary statistics for the pre- and post-PPP periods, before and after PSM. Table 1 compares all PPP recipients to the full sample of non-recipients. For each variable, I show the pre- and post-PPP means, standard deviations, and the DID between treated and untreated firms. Outcomes include employment, log(employment), the Financial Stress and Commercial Credit rankings, while controls include log(sales), import and export status, and the stratification variables (ethnic minority ownership, public vs. private status, and 7(a) history).

Before matching, PPP recipients are systematically different from non-recipients. They are larger (higher employment and log(employment)), have higher average sales, and display somewhat better pre-PPP Financial Stress and Commercial Credit rankings. Several of these differences are statistically significant. This pattern is consistent with selection into PPP on observables: firms that were bigger, more established, and somewhat less risky prior to the program were more likely to obtain loans. At the same time, differences in key stratification variables such as ethnic minority ownership, public status, and 7(a) history are small in magnitude, suggesting that PPP take-up was not driven exclusively by these categorical characteristics.

Table 2 repeats the same comparison using the matched sample of non-recipients and the same set of PPP recipients. After PSM, the distributions of core covariates are much more similar across treated and control firms. Differences in log(sales), import and export status, and the stratification variables are small and statistically insignificant, and pre-PPP differences in outcomes are also substantially reduced. In other words, the matching procedure delivers a control group that closely resembles PPP recipients along observed dimensions.

Compared with Table 1, Table 2 shows that PSM helps the covariates become much better balanced, as virtually all of them exhibit insignificant differences between treated and control firms. This balancing on observables strengthens the credibility of the subsequent DID estimates and suggests that, conditional on the rich set of matched covariates, unobserved characteristics are likely to be more homogeneous across PPP and non-PPP firms in the matched sample than in the raw data.

**B. Employment**

Table 3 Panel A reports the baseline effects of PPP loans on employment. In the pooled specification, PPP recipients have about 0.068% higher employment than matched non-recipients. The effect is small in percentage terms but precisely estimated and applies to a very large treated population. Splitting by minority ownership, the estimates remain positive for both groups: employment is about 0.062% higher for minority-owned firms and 0.047% higher for non-minority firms relative to their matched controls.

Model (2) replaces the PPP dummy with loan duration (months between April and August 2020). Each additional month of PPP coverage is associated with roughly 0.015% higher employment, implying a cumulative effect of about 0.075% over five months, very close to the baseline 0.068% treatment effect. Duration gradients for minority and non-minority firms are similar.

Table 4 examines heterogeneity in employment effects. Panel A stratifies firms by pre-PPP size and shows a non-linear ("parabolic") pattern. Medium-small firms (10–99 employees) and medium-sized firms (100–499 employees) exhibit the largest gains, about 0.274% and 0.276% higher employment, respectively, relative to non-recipients. Micro firms (1–4 employees) show smaller gains (0.051%), and large firms have more modest increases (0.114%). In the duration specification, medium-small firms gain about 0.057% per month (≈ 0.285% over five months), while micro firms gain around 0.012% per month; effects for large firms are small and often statistically insignificant.

Panels B and C stratify by pre-PPP financial condition. Panel B uses the Financial Stress Score (FSS). Firms with the highest failure risk display employment gains of about 0.195%, whereas low-risk firms do not exhibit statistically significant effects. Panel C uses the Commercial Credit Score (CCS). Firms with CCS = 2 (moderately low delinquency risk) gain about 0.083% in employment; firms with CCS = 1 and 5 gain only 0.009% and 0.028%, respectively. In the duration specification, CCS = 2 firms gain about 0.018% per month (≈ 0.09% over five months), while CCS = 1 firms gain roughly 0.002% per month. Overall, employment effects concentrate among firms of intermediate size and intermediate pre-PPP risk.

**C. Failure Risk - Financial Stress**

Table 3 Panel B reports the impact of PPP on financial stress rankings, measured by the Failure Score percentile (higher percentiles = lower failure risk). In the pooled sample, PPP recipients improve their Failure Score percentile by about 1.197 points relative to matched non-recipients. Minority-owned firms improve by about 0.095 points, while non-minority firms improve by about 1.256 points.

Model (2) considers loan duration. Each additional month of PPP coverage increases the Failure Score percentile by roughly 0.173 points, for a total of about 0.865 points over five months. Minority-owned firms gain about 0.025 points in total, whereas non-minority firms gain about 0.890 points.

Table 5 explores heterogeneity in financial-stress effects. Panel A stratifies by firm size. Micro firms (1–4 employees) and medium-small firms (10–99 employees) have the largest improvements, about 1.252 and 1.360 percentile points, respectively. Small firms (5–9 employees) gain about 0.124 points, and large firms show no significant change. In the duration specification, micro and medium-small firms gain roughly 0.267 and 0.285 points per additional month of PPP coverage, while small and large firms see minimal or insignificant changes.

Panel B stratifies by pre-PPP FSS risk class. Firms with medium initial stress levels exhibit the largest improvement, about 3.429 percentile points, while low-stress firms gain about 0.632 points, and the highest-stress group improves less. With loan duration, medium-stress firms gain about 0.771 points per month ($\approx$ 3.855 over five months), compared with roughly 0.135 points in total for low-stress firms. PPP thus has the strongest impact on failure-risk rankings for firms with intermediate levels of pre-PPP financial stress.

### D. Delinquency Risk - Commercial Credit

Table 3 Panel C evaluates PPP's effect on commercial credit, using the Delinquency Predictor Score percentile (Commercial Credit Score, CCS; higher percentiles = lower delinquency risk). PPP recipients improve their delinquency percentile by about 3.245 points relative to matched non-recipients. Minority-owned recipients gain about 1.795 points, while non-minority firms gain about 3.307 points.

Model (2) shows that loan duration is also important. Each additional month of PPP coverage is associated with about 0.569 percentile points higher CCS, i.e., about 2.845 points over five months. Minority-owned firms gain about 0.335 points per month ($\approx$ 1.675 total), and non-minority firms about 0.575 points per month ($\approx$ 2.875 total).

Table 6 documents heterogeneity in commercial-credit effects. Panel A stratifies by firm size. Micro firms (1–4 employees) experience the largest improvement, about 3.424 CCS percentile points, while medium-sized firms (100–499 employees) improve by about 1.454 points. For large firms, the estimates are small and not statistically significant. In the duration specification, micro firms gain about 0.729 points per month ($\approx$ 3.645 over five months), medium-sized firms about 0.318 points per month, and large firms show a small, non-significant decline ($\approx$ –0.091 per month).

Panel B stratifies by pre-PPP commercial credit risk. Firms with medium initial delinquency risk show the largest gains, about 11.187 percentile points, whereas low-risk firms see only about 0.083 points of improvement. With loan duration, medium-risk firms gain about 2.622 points per additional month ($\approx$ 13.110 over five months), while low-risk firms gain only about 0.018 points in total. PPP therefore generates substantial improvements in commercial credit conditions, especially for smaller firms and those with intermediate levels of pre-PPP credit risk.

### E. Heterogeneity by 7(a) Loan and Industry

Appendix Table C.1 examines heterogeneity by prior SBA 7(a) borrowing and by broad industry groups. Columns (1)–(2) show that PPP raises log employment by about 1.8–2.0% for both firms with and without 7(a) histories, relative to matched non-recipients. Columns (3)–(4) indicate that Failure Score percentiles increase by about 0.9–1.1 points in both groups.

The contrast is sharper for commercial credit. Columns (5)–(6) show that PPP improves CCS percentiles by about 1.3 points for firms with 7(a) histories, but by more than 3.2 points for firms without prior SBA loans. Panel B further shows that longer loan duration is associated with additional gains in all three outcomes, especially for non-7(a) firms.

Across industry groups, PPP effects are strongest in sectors most exposed to COVID-19 shocks, such as retail and transportation, accommodation and food services, entertainment, and selected public administration and social services, where estimated employment, Failure Score, and CCS gains are larger and more precisely estimated. In less directly affected sectors, the corresponding effects are smaller and often statistically weaker.

## Section VI. Discussion and Implications

### A. Main Impacts and Rebalancing PPP Relative to Prior Evidence

This study reassesses the early impact of the Paycheck Protection Program using a near-universe panel of U.S. firms and three outcomes: employment, financial stress, and commercial credit risk. Baseline estimates show that PPP receipt raised employment by about 0.068 percent relative to matched non-recipients over April–September 2020. Effects on forward-looking risk measures are larger: recipients improve their Financial Stress percentile ranking by roughly 1.2 points and their Commercial Credit percentile ranking by about 3.2 points. Loan duration matters on all three margins, with longer coverage associated with incremental gains in employment and especially in risk scores. PPP thus appears to have operated at least as much as a balance-sheet and credit-risk intervention as a pure jobs-preservation program.

Relative to the existing PPP literature, the employment effect here is small. Using Dun & Bradstreet data for firms with loans above $150,000 and early PPP records, Hubbard and Strain (2021) estimate employment gains of about 0.903 percent for PPP-eligible firms. Autor et al. (2021a, 2021b), using ADP payroll microdata for about three million firms of all sizes, find that PPP increased employment by 4–10 percent in the short run and 0–6 percent longer-run for larger firms, with roughly twice those magnitudes for firms with 1–49 employees. Chetty et al. (2023) report employment responses around 2 percent for small businesses. By comparison, this study's 0.068 percent effect is an average impact for PPP recipients relative to matched non-recipients over the same early period.

Several design and data differences help explain the divergence. Much prior work constructs intent-to-treat estimates around the 500-employee cutoff or access to particular intermediaries (for example, large payroll providers). In practice, the 500-employee rule was porous: many firms above 500 employees remained eligible under industry-specific rules, and many smaller firms did not receive loans (Humphries, Neilson, and Ulyssea (2020); Granja et al. (2020)). Here, I use a panel of about 30.4 million firms tracked by Dun & Bradstreet, merged with full SBA PPP and 7(a) records, and estimate treatment-on-the-treated effects of actually receiving and holding PPP loans relative to matched non-recipients.

The empirical strategy also differs. Hubbard and Strain (2021), Autor et al. (2021a, 2021b), and Chetty et al. (2023) mainly exploit repeated cross-sections or designs around eligibility rules. I implement a PSM–DID framework on a very large panel, matching on rich pre-treatment characteristics and differencing over time. Coverage and outcome scaling are also different: earlier studies focus on intensive margins of payroll employment for firms strongly represented in payroll or tax data, or near the eligibility threshold. This study instead covers almost the entire population of firms with measurable credit demand, with micro-businesses (1–4 employees) constituting the majority and receiving mostly loans below $150,000. The 0.068 percent effect should therefore be interpreted as a population-weighted average across a broad and heterogeneous treated group, not just the higher-impact segments emphasized earlier.

Risk and credit results speak to a dimension that early work only partially addressed. Hubbard and Strain (2021) report modest improvements in PAYDEX trade-payment scores, but PAYDEX is observed for only a minority of firms, raising selection concerns. Joaquim and Wang (2022) use Dun & Bradstreet's Failure Stress Score to show improved solvency risk for recipients. Building on these contributions, I follow almost the full universe of firms with credit demand and jointly track employment, Failure Score percentiles and Commercial Credit Score (CCS) percentiles, both widely used by lenders and public agencies. The evidence indicates that PPP loans were associated with small average employment gains but more substantial improvements in failure and delinquency risk. From the vantage point of late 2025, this rebalances the PPP narrative: earlier work emphasised jobs saved and cost per job; the near-universe credit-bureau view suggests that a central contribution of PPP was to reduce predicted business failure and severe delinquency across a very broad base, while only modestly raising employment at the average treated firm.

## B. Who Benefited, and Along Which Margins?

The heterogeneity analysis shows that these averages conceal sharp differences. A first axis is firm size. Firms with 10–99 and 100–499 employees experience the largest gains in employment and financial-stress rankings, while micro firms with 1–4 workers and near-threshold larger firms see smaller or statistically weaker improvements. This matches evidence that small and medium-sized enterprises (SMEs) have enough organisational capacity to absorb and deploy external funding, yet remain heavily dependent on bank credit and cash flow (Beck, Demirgüç-Kunt, and Maksimovic (2005); Acs and Mueller (2008); Ayyagari, Demirgüç-Kunt, and Maksimovic (2008)). Micro firms often lack formal financial infrastructure; larger firms have diversified financing and buffers, so PPP is one stabilising tool among many rather than decisive.

A second axis is pre-PPP risk. Stratifying by pre-PPP Failure Score and CCS reveals an inverted-U pattern: firms in the middle of the risk distribution exhibit the largest improvements in both stress and credit rankings, while very low-risk firms change little and the most distressed firms improve only modestly. This aligns with work on credit guarantees and bank support that finds the strongest responses among borrowers that are financially constrained but still viable (Jiménez, Ongena, Peydró, and Saurina (2012); Berger and Roman (2017)). For very safe firms, PPP mainly augments existing liquidity buffers. For highly stressed firms, pre-existing leverage, arrears, and structural problems limit the extent to which a temporary, partially forgivable loan can durably reduce default or delinquency risk. Ex post, these results suggest that targeting by risk band, not just by size or revenue loss, could materially improve the cost-effectiveness of similar interventions.

Distributional margins are also important. Conditional on receiving a PPP loan, minority-owned firms show slightly larger employment gains than non-minority firms but smaller improvements in financial-stress and credit rankings. This is consistent with PPP funds being used primarily to maintain payroll in minority-owned businesses without fully closing underlying gaps in balance-sheet strength and perceived creditworthiness. The broader literature documents greater barriers and later or smaller PPP access for minority-owned small businesses (Humphries, Neilson, and Ulyssea (2020); Granja et al. (2020); Fairlie and Fossen (2021); Atkins, Cook, and Seamans (2021)). Emergency loan programs can therefore offset some labour-market impacts for historically disadvantaged firms, but do not by themselves eliminate structural inequities in credit markets.

Prior SBA relationships provide another lens. Firms with a pre-PPP history of 7(a) borrowing benefit in terms of employment and risk scores, consistent with relationship lenders helping viable clients navigate PPP quickly. At the same time, firms without 7(a) histories experience substantially larger improvements in CCS, suggesting that PPP also acted as a one-time extension of the SBA frontier into previously under-served segments. During the crisis, reliance on existing banks and SBA channels raised concerns that relationship borrowers would be favoured (Erel and Liebersohn (2020); Balyuk, Prabhala, and Puri (2021)). The results here show a more nuanced picture: relational borrowers gained through faster access and employment stabilisation, while previously unconnected firms used PPP to improve documented credit profiles, potentially easing future finance.

Industry patterns reinforce the value of combining sectoral information with firm-level risk scores. Employment responses are strongest in manufacturing and high-contact services, while improvements in financial-stress and credit scores are especially pronounced in sectors most exposed to shutdowns and payment delays, such as retail, food and accommodation, and entertainment. The positive association between loan duration and outcomes across these sectors underscores the importance of the liquidity horizon: firms in industries with slower demand recoveries or more rigid cost structures gain disproportionately from longer maturities and extended forgiveness windows.

Finally, the 2020–2021 PPP wave marks a shift toward using high-frequency, firm-level data in policy evaluation. This study shows that commercial credit-bureau data can be integrated with public loan records to track not only jobs and revenues but also solvency and credit conditions for almost the full firm population. For future crises, this points to using existing risk scores and firm registries not just for ex post evaluation but for ex ante design, helping to target support where marginal dollars are most productive. From the vantage point of late 2025, PPP appears as an extremely large, rapidly deployed intervention that modestly raised employment but more clearly improved failure and delinquency risk for small–medium, medium-risk, and previously less-connected firms.

## C. Policy Implications

The results have several implications for the design of PPP-type interventions and the broader architecture of emergency business support.

First, the sharp heterogeneity by firm size and pre-crisis risk suggests that uniform subsidies are a blunt instrument. The strongest effects on employment and risk scores are concentrated among small–medium firms with 10–499 employees and intermediate levels of financial stress and credit risk, while micro-enterprises and highly distressed firms benefit less. This pattern indicates that, in future crises, key parameters, loan caps, forgiveness rates, co-financing requirements, should be calibrated by size and risk band, not size alone. Even coarse segmentation (micro vs. small–medium vs. near-threshold large firms, and low-/medium-/high-risk borrowers based on pre-crisis bureau scores) would push scarce fiscal resources toward segments where marginal dollars generate the largest real effects.

Second, the evidence on loan duration and timing highlights the importance of the liquidity horizon, not just initial disbursement. Longer PPP holding durations are associated with systematically better outcomes on all three margins, especially financial stress and commercial credit. Combined with the information frictions documented by Humphries, Neilson, and Ulyssea (2020) and others, this yields two design lessons. In the acute phase, policymakers should invest in early, broad information outreach to mitigate "first-come, first-served" bias and ensure that viable but less-connected firms can access support quickly. As the shock evolves, maturity structures and forgiveness rules should be differentiated across sectors and risk profiles, with longer horizons for industries facing slower demand recovery or more rigid cost structures. Crisis lending should be seen as time-contingent insurance, not a one-off transfer.

Third, the findings on minority-owned firms and on 7(a) relationships underscore that emergency programs do not by themselves close structural credit gaps, but they can shift the starting point for future policy. Conditional on receiving PPP, minority-owned firms achieve at least comparable employment gains but smaller improvements in financial-stress and credit rankings than non-minority firms, consistent with PPP helping them hold payroll without fully offsetting longstanding disadvantages in access to capital and perceived risk. At the same time, firms without prior 7(a) histories show particularly large gains in Commercial Credit scores, indicating that PPP served as a first formal credit relationship for many previously unbanked or under-banked businesses. This points to a two-step approach: in a crisis, incorporate explicit mechanisms, reserved volumes, dedicated fintech channels, targeted outreach, to prevent minority-owned and historically underserved firms from being systematically late to the queue; after the crisis, treat the new relationships and data generated by PPP as a platform for durable inclusion, for example by streamlining transitions into regular 7(a) products or complementary technical assistance.

Fourth, linking PPP records with near-universal credit-bureau data illustrates how existing information systems can be used not only for ex post evaluation but for ex ante design. PPP in 2020 was built under extreme time pressure, and the debate emphasized a trade-off between speed and targeting (Hubbard and Strain (2020); Granja et al. (2020)). By 2025, it is clear that high-frequency administrative and commercial data can be linked at scale with modest lags. The same Failure Score and Commercial Credit Score measures used here to assess outcomes could, in principle, have been used ex ante to vary loan size, guarantee intensity, or forgiveness terms by risk band, or to flag firms whose pre-crisis stress levels made them unlikely to be stabilized by small, short-term loans alone. The lesson is not that every emergency program should be fully "credit-scored," but that existing risk information can improve targeting at low marginal cost once legal and technical frameworks are in place.

Finally, viewed from late 2025, PPP should be understood as part of a broader shock-response infrastructure rather than a one-off experiment. This study shows that PPP modestly raised employment for the average treated firm but more visibly improved failure and delinquency risk across a broad base of small and medium-sized enterprises, with effects varying systematically by size, risk, ownership, banking relationships, and sector. For future crises, whether pandemics, natural disasters, or abrupt credit contractions, the main lessons are that speed and simplicity remain crucial, but they need not preclude using readily available information on firms' size, risk, and sectoral exposure. Programs that build on existing administrative and commercial data, that recognize where marginal dollars are most productive, and that deliberately extend the frontier of financial inclusion are more likely to deliver both short-run stabilization and a more resilient firm sector in the long run.

# Tables and Figures

Table 1. Descriptive Statistics for PPP Recipients and Non-Recipients for the Complete Sample

|  | Pre-PPP Mar 2020 | | Post-PPP Sep 2020 | | ΔΔ |
|---|---|---|---|---|---|
|  | Non-Recipients | PPP Recipients | Non-Recipients | PPP Recipients |  |
| ***Dependent Variables***  | | | | | |
| Employees | 7.569 | 9.983 | 7.289 | 9.942 | 0.239*** |
|  | (148.619) | (62.431) | (111.469) | (58.116) | (0.057) |
| log (Employees) | 1.473 | 1.734 | 1.443 | 1.728 | 0.024*** |
|  | (0.808) | (0.893) | (0.809) | (0.896) | (0.001) |
| Financial Stress Ranking | 37.673 | 50.898 | 37.985 | 52.195 | 0.985*** |
|  | (24.288) | (27.286) | (24.050) | (26.830) | (0.021) |
| Commercial Credit Ranking | 44.633 | 60.996 | 44.130 | 63.433 | 2.939*** |
|  | (28.994) | (32.165) | (29.187) | (31.999) | (0.025) |
| ***Control Variables***  | | | | | |
| Annual Sales Amount ($) | 1,910,854 | 4,163,279 | 1,924,637 | 4,182,503 |  |
|  | (172,311,560) | (336,555,100) | (174,147,600) | (367,743,600) |  |
| log (Sales) | 10.084 | 12.167 | 10.080 | 12.175 | 0.012*** |
|  | (3.869) | (2.319) | (3.873) | (2.323) | (0.002) |
| Import | 0.005 | 0.003 | 0.005 | 0.003 | 0.000 |
|  | (0.067) | (0.055) | (0.067) | (0.055) | (0.000) |
| Export | 0.002 | 0.001 | 0.002 | 0.001 | 0.000*** |
|  | (0.062) | (0.001) | (0.062) | (0.001) | (0.000) |
| ***Stratification Variables***  | | | | | |
| 7(a) Experience | 0.008 | 0.032 | 0.007 | 0.032 | 0.001 |
|  | (0.089) | (0.175) | (0.086) | (0.175) | (0.000) |
| Minority Status | 0.003 | 0.026 | 0.003 | 0.025 | -0.001 |
|  | (0.057) | (0.160) | (0.057) | (0.156) | (0.000) |
| Public Status | 0.009 | 0.001 | 0.009 | 0.001 | 0.000 |
|  | (0.096) | (0.026) | (0.095) | (0.026) | (0.000) |

*Notes:* This table reports summary statistics for the dependent variables, controls in specification (1), and stratification variables for the full sample, comparing PPP recipients (treatment, 3,723,513 firms) and non-recipients (control, 26,646,677 firms) in March 2020 (pre-PPP) and September 2020 (post-PPP). Ethnic minority equals 1 if the owner is Hispanic, Black, African American, American Indian, Asian, Native Hawaiian, Pacific Islander, Alaskan Native, or multiple races. Public indicates whether the firm is publicly listed. The last two columns report DID t-tests. ***, **, and * denote significance at the 1%, 5%, and 10% levels, respectively.
*Source:* Author's calculations based on D&B data.

Table 2. Descriptive Statistics for PPP Recipients and Non-Recipients for the Matched Sample

|  | Pre-PPP Mar 2020 | | Post-PPP Sep 2020 | | ΔΔ |
| --- | --- | --- | --- | --- | --- |
|  | Non-Recipients | PPP Recipients | Non-Recipients | PPP Recipients | |
| *Dependent Variables* | | | | | |
| Employees | 8.645 | 9.983 | 8.751 | 9.942 | 0.104*** |
|  | (97.781) | (62.431) | (94.424) | (58.116) | (0.022) |
| log (Employees) | 1.271 | 1.734 | 1.259 | 1.728 | 0.012*** |
|  | (0.844) | (0.893) | (0.844) | (0.896) | (0.000) |
| Financial Stress Ranking | 46.020 | 50.898 | 47.115 | 52.195 | 0.621*** |
|  | (26.924) | (27.286) | (26.752) | (26.830) | (0.010) |
| Commercial Credit Ranking | 51.817 | 60.996 | 53.841 | 63.433 | 1.375*** |
|  | (31.233) | (32.165) | (31.707) | (31.999) | (0.009) |
| *Control Variables* | | | | | |
| Annual Sales Amount ($) | 1,886,277 | 4,163,279 | 1,900,926 | 4,182,503 | |
|  | (171,110,800) | (336,555,100) | (172,424,400) | (367,743,600) | |
| log (Sales) | 11.031 | 12.167 | 11.029 | 12.175 | 0.005 |
|  | (3.388) | (2.319) | (3.401) | (2.323) | (0.000) |
| Import | 0.002 | 0.003 | 0.002 | 0.003 | 0.000 |
|  | (0.049) | (0.055) | (0.050) | (0.055) | (0.000) |
| Export | 0.001 | 0.001 | 0.001 | 0.001 | 0.000 |
|  | (0.001) | (0.001) | (0.001) | (0.001) | (0.000) |
| *Stratification Variables* | | | | | |
| 7(a) Experience | 0.004 | 0.032 | 0.004 | 0.032 | 0.000 |
|  | (0.058) | (0.175) | (0.056) | (0.175) | (0.000) |
| Minority Status | 0.013 | 0.026 | 0.013 | 0.025 | 0.000 |
|  | (0.114) | (0.160) | (0.107) | (0.156) | (0.000) |
| Public Status | 0.003 | 0.001 | 0.003 | 0.001 | 0.000 |
|  | (0.050) | (0.026) | (0.051) | (0.026) | (0.000) |

*Notes:* This table reports summary statistics for the dependent variables, controls in specification (1), and stratification variables on the matched sample, comparing PPP recipients (treatment, 3,723,513 firms) and matched non-recipients (control, 3,723,513 firms) in March 2020 (pre-PPP) and September 2020 (post-PPP). Ethnic minority equals 1 if the owner is Hispanic, Black, African American, American Indian, Asian, Native Hawaiian, Pacific Islander, Alaskan Native, or multiple races. Public indicates whether the firm is publicly listed. The last two columns report DID t-tests. ***, **, and * denote significance at the 1%, 5%, and 10% levels, respectively.
*Source:* Author's calculations based on D&B data.

Table 3. Effects of PPP Loans on log(employment)

|  | (1) Full Sample | (2) Minority | (3) Non-Minority |
|---|---|---|---|
| Mean of Duration | 0.345 | 0.673 | 0.267 |
| Std of Duration | (1.178) | (1.612) | (1.074) |
| *Panel A* | | | |
| Mean of log(employment) | 2.087 | 2.254 | 2.085 |
| Std of log(employment) | (0.989) | (1.787) | (0.987) |
| **Model (1)** | | | |
| $PPP_i * Post$ | 0.068*** | 0.062*** | 0.047*** |
|  | (0.002) | (0.002) | (0.002) |
| **Model (2)** | | | |
| $Duration_{it}$ | 0.015*** | 0.014*** | 0.015*** |
|  | (0.002) | (0.002) | (0.002) |
| *Panel B* | | | |
| Mean of Financial Stress Rk. | 39.511 | 46.751 | 39.439 |
| Std err of Financial Stress Rk. | (24.959) | (26.851) | (24.929) |
| **Model (1)** | | | |
| $PPP_i * Post$ | 1.197*** | 0.095*** | 1.256*** |
|  | (0.002) | (0.002) | (0.002) |
| **Model (2)** | | | |
| $Duration_{it}$ | 0.173*** | 0.005 | 0.178*** |
|  | (0.002) | (0.002) | (0.002) |
| *Panel C* | | | |
| Mean of Commercial Credit Rk. | 46.556 | 51.618 | 46.505 |
| Std err of Commercial Credit Rk. | (30.055) | (33.007) | (30.020) |
| **Model (1)** | | | |
| $PPP_i * Post$ | 3.245*** | 1.795*** | 3.307*** |
|  | (0.001) | (0.062) | (0.039) |
| **Model (2)** | | | |
| $Duration_{it}$ | 0.569 | 0.335*** | 0.575*** |
|  | (0.002) | (0.014) | (0.002) |
| Observations | 7,447,026 | 465,440 | 6,981,586 |

*Notes:* Models (1) and (2) are defined in the text. All specifications include state-by-month and industry-by-month fixed effects. The time dummy Post equals 0 in March 2020 and 1 in September 2020. Columns (2) and (3) report estimates stratified by owner minority status. Minority includes firms whose owner identifies as Hispanic, Black, African American, American Indian, Asian, Native Hawaiian, Pacific Islander, Alaskan Native, or multiple races; non-Minority includes all others. Standard errors are in brackets. ***, **, and * indicate significance at the 1%, 5%, and 10% levels, respectively.
*Source:* Author's calculations based on D&B data.

Table 4. Heterogeneity of the Effects of PPP Loans on log(employment)

| *Panel A. By Size* | Micro<br>Size 1-4 | Small<br>Size 5-9 | S-M<br>Size 10-99 | Med<br>Size 100-499 | Large<br>Size > 500 |
|---|---|---|---|---|---|
| Mean of log(emp) | 1.470 | 2.999 | 5.151 | 7.166 | 7.431 |
| Std err of log(emp) | (0.448) | (0.409) | (0.707) | (0.893) | (1.350) |
| **Model (1)** | | | | | |
| $PPP_i * Post$ | 0.051*** | 0.127*** | 0.274*** | 0.276*** | 0.114*** |
|  | (0.001) | (0.001) | (0.001) | (0.005) | (0.025) |
| **Model (2)** | | | | | |
| $Duration_{it}$ | 0.012*** | 0.029*** | 0.057*** | 0.029*** | -0.011*** |
|  | (0.000) | (0.000) | (0.000) | (0.006) | (0.002) |
| Observations | 4,982,451 | 2,106,311 | 276,869 | 75,075 | 6,320 |
| *Panel B. By pre-FSS* | 1<br>Low Failure<br>Risk | 2 | 3 | 4 | 5<br>High Failure<br>Risk |
| Mean of log(emp) | 2.064 | 1.552 | 1.440 | 1.480 | 1.657 |
| Std err of log(emp) | (1.252) | (0.892) | (0.830) | (0.765) | (0.958) |
| **Model (1)** | | | | | |
| $PPP_i * Post$ | 0.000 | 0.028*** | 0.049*** | 0.037*** | 0.195*** |
|  | (0.000) | (0.000) | (0.001) | (0.001) | (0.011) |
| **Model (2)** | | | | | |
| $Duration_{it}$ | 0.000 | 0.006*** | 0.011*** | 0.009*** | 0.044*** |
|  | (0.000) | (0.000) | (0.000) | (0.000) | 0.000 |
| Observations | 128,086 | 1,166,522 | 2,377,180 | 3697830 | 81408 |
| *Panel C. By pre-CCS* | 1<br>Low Delinq.<br>Risk | 2 | 3 | 4 | 5<br>High Delinq.<br>Risk |
| Mean of log(emp) | 2.331 | 2.267 | 1.918 | 1.897 | 1.341 |
| Std err of log(emp) | (1.076) | (1.156) | (0.842) | (0.844) | (0.645) |
| **Model (1)** | | | | | |
| $PPP_i * Post$ | 0.009*** | 0.083*** | 0.033*** | 0.040*** | 0.028*** |
|  | (0.002) | (0.001) | (0.001) | (0.001) | (0.014) |
| **Model (2)** | | | | | |
| $Duration_{it}$ | 0.002*** | 0.018*** | 0.007*** | 0.009*** | 0.008*** |
|  | (0.000) | (0.000) | (0.000) | (0.000) | (0.000) |
| Observations | 488,695 | 1,637,705 | 2,314,909 | 2,970,251 | 35,466 |

*Notes:* Models (1) and (2) are defined in the text. All specifications include state-by-month and industry-by-month fixed effects. The time dummy Post equals 0 in March 2020 and 1 in September 2020. Panel A stratifies firms by pre-PPP size, measured as average employment over the 12 months before PPP. Panel B stratifies by pre-PPP financial stress scores. Panel C stratifies by pre-PPP commercial credit scores. Standard errors are in brackets. ***, **, and * indicate significance at the 1%, 5%, and 10% levels, respectively.
*Source:* Author's calculations based on D&B data.

Table 5. Heterogeneity of the Effects of PPP Loans on Financial Stress Ranking Percentile

| *Panel A. By Size* | Micro<br>Size 1-4 | Small<br>Size 5-9 | S-M<br>Size 10-99 | Med<br>Size 100-499 | Large<br>Size > 500 |
|---|---|---|---|---|---|
| Mean of Fin Stress Rk. | 41.112 | 34.324 | 49.359 | 47.192 | 45.058 |
| Std err of Fin Stress Rk. | (24.616) | (23.924) | (30.100) | (30.448) | (28.046) |
| **Model (1)** | | | | | |
| $PPP_i * Post$ | 1.252*** | 0.124*** | 1.360*** | 0.780*** | 0.264 |
| | (0.036) | (0.034) | (0.022) | (0.058) | (0.203) |
| **Model (2)** | | | | | |
| $Duration_{it}$ | 0.267*** | 0.003 | 0.285*** | 0.170*** | 0.076 |
| | (0.001) | (0.002) | (0.005) | (0.012) | (0.047) |
| Observations | 4,982,451 | 2,106,311 | 276,869 | 75,075 | 6,320 |
| *Panel B. By pre FSS* | 1<br>Low Failure Risk | 2 | 3 | 4 | 5<br>High Failure Risk |
| Mean of Fin Stress Rk. | 95.602 | 77.646 | 47.752 | 21.083 | 1.671 |
| Std err of Fin Stress Rk. | (6.956) | (11.813) | (12.764) | (11.019) | (4.393) |
| **Model (1)** | | | | | |
| $PPP_i * Post$ | 0.632*** | 1.183*** | 2.345*** | 3.429*** | 1.218*** |
| | (0.021) | (0.030) | (0.036) | (0.039) | (0.065) |
| **Model (2)** | | | | | |
| $Duration_{it}$ | 0.135*** | 0.258*** | 0.511*** | 0.771*** | 0.225*** |
| | (0.004) | (0.002) | (0.002) | (0.002) | (0.005) |
| Observations | 128,086 | 1,166,522 | 2,377,180 | 3,697,830 | 81,408 |

*Notes:* DID model (1) uses PPP recipients as the treatment group and PSM-matched non-recipients as the control group. All specifications include state-by-month and industry-by-month fixed effects. The time dummy *Post* equals 0 in March 2020 and 1 in September 2020. The table contains three panels: Panel A stratifies firms by pre-PPP employee count, Panel B by pre-PPP financial stress score, and Panel C by pre-PPP commercial credit score. In each panel, Model (1) reports the average PPP effect and Model (2) the effect of loan duration, measured as the number of months the loan is held through September 2020 (1–5). Standard errors are in brackets. ***, **, and * indicate significance at the 1%, 5%, and 10% levels, respectively.
*Source:* Author's calculations based on D&B data.

Table 6. Heterogeneity of the Effects of PPP Loans on Commercial Credit Ranking Percentile

| *Panel A. By Size* | Micro<br>Size 1-4 | Small<br>Size 5-9 | S-M<br>Size 10-99 | Med<br>Size 100-499 | Large<br>Size > 500 |
|---|---|---|---|---|---|
| Mean of Com Credit Rk. | 46.958 | 42.966 | 64.291 | 62.477 | 54.132 |
| Std err of Com Credit Rk. | (29.986) | (28.935) | (31.915) | (30.941) | (32.200) |
| **Model (1)** | | | | | |
| $PPP_i * Post$ | 3.424*** | 2.107*** | 2.536*** | 1.454*** | -0.677* |
|  | (0.052) | (0.052) | (0.041) | (0.115) | (0.887) |
| **Model (2)** | | | | | |
| $Duration_{it}$ | 0.729*** | 0.414*** | 0.530*** | 0.318*** | -0.091 |
|  | (0.002) | (0.003) | (0.009) | (0.024) | (0.092) |
| Observations | 4,982,451 | 2,106,311 | 276,869 | 75,075 | 6,320 |
| *Panel B. By pre CCS* | 1<br>Low Delinq.<br>Risk | 2 | 3 | 4 | 5<br>High Delinq.<br>Risk |
| Mean of Com Credit Rk. | 94.885 | 79.400 | 48.834 | 19.259 | 1.398 |
| Std err of Com Credit Rk. | (11.603) | (16.395) | (14.469) | (13.362) | (2.596) |
| **Model (1)** | | | | | |
| $PPP_i * Post$ | 0.083*** | 2.105*** | 6.029*** | 11.187*** | 1.341*** |
|  | (0.017) | (0.051) | (0.052) | (0.059) | (0.025) |
| **Model (2)** | | | | | |
| $Duration_{it}$ | 0.018*** | 0.457*** | 1.344*** | 2.622*** | 0.324*** |
|  | (0.004) | (0.003) | (0.003) | (0.003) | (0.006) |
| Observations | 488,695 | 1,637,705 | 2,314,909 | 2,970,251 | 35,466 |

*Notes:* DID model (1) uses PPP recipients as the treatment group and PSM-matched non-recipients as the control group. All specifications include state-by-month and industry-by-month fixed effects. The time dummy *Post* equals 0 in March 2020 and 1 in September 2020. The table has three panels: Panel A stratifies by pre-PPP employee count, Panel B by pre-PPP financial stress score, and Panel C by pre-PPP commercial credit score. In each panel, Model (1) reports the average PPP effect and Model (2) the effect of loan duration, defined as the number of months the loan is held through September 2020 (1–5). Standard errors are in brackets. ***, **, and * indicate significance at the 1%, 5%, and 10% levels, respectively.
*Source:* Author's calculations based on D&B data.

Figure 1. Pre- and Post-PPP Employment Trends by Firm Size and Loan Approval Status

(a) Employment count of all businesses of different sizes during the PPP impact period.

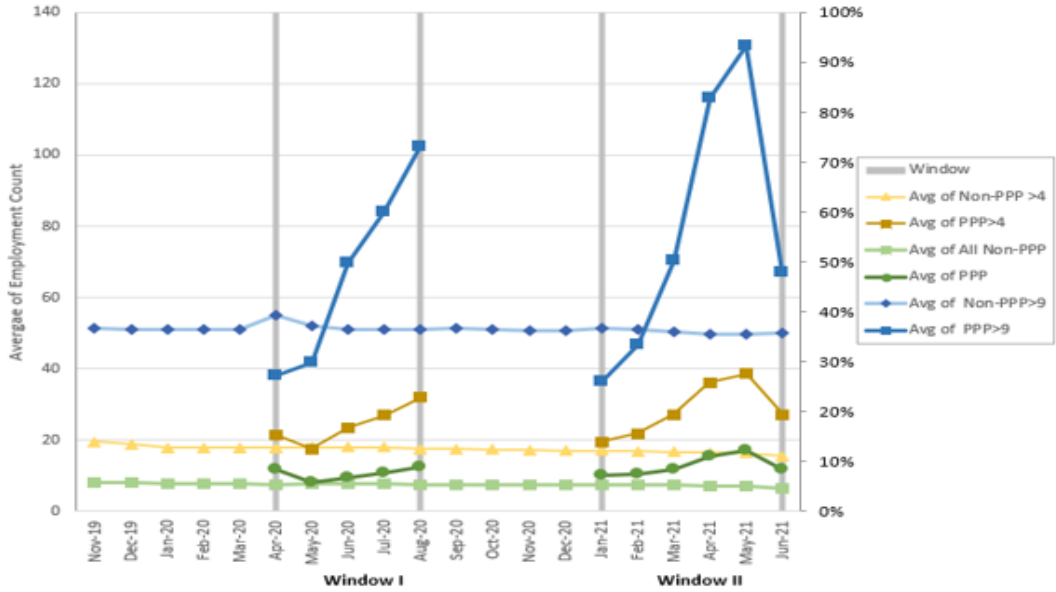

(b) Employment count of larger businesses during the PPP impact period.

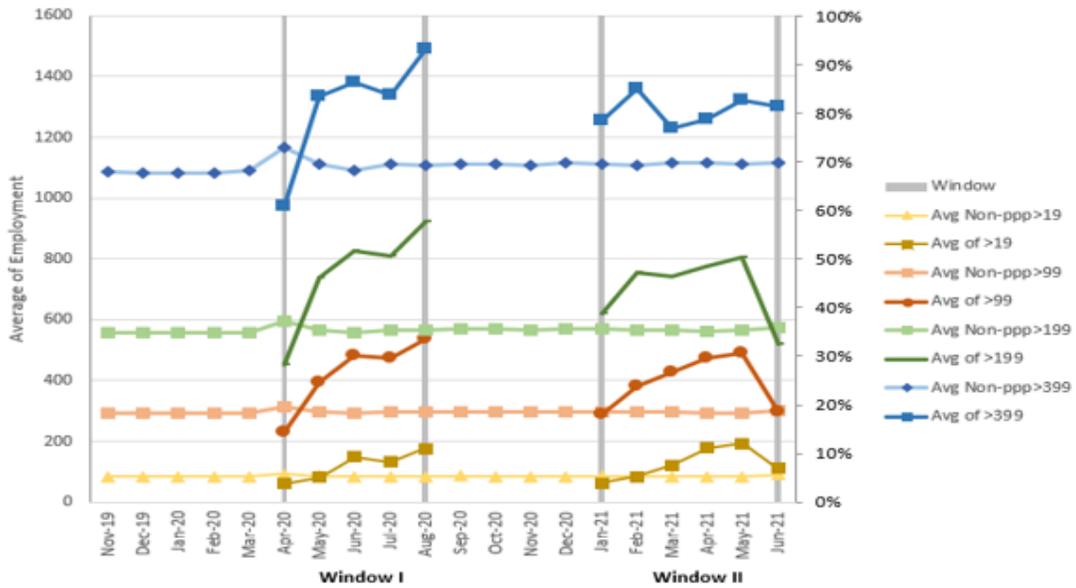

*Notes:* Panel (a) plots monthly average establishment-level employment for all firms, by PPP approval status and firm-size groups, from November 2019 to June 2021. Panel (b) repeats the exercise for larger firms only. Solid lines report mean employment (left axis), while dashed lines report the share of PPP recipients within each size group (right axis). Vertical lines indicate the beginning of the first PPP window (April 2020) and the second PPP window (January 2021). "PPP" firms are establishments that received at least one PPP loan by June 2021; "Non-PPP" firms are observably similar establishments that did not receive PPP funds over the sample period.
*Source:* Author's calculations based on Dun & Bradstreet datasets described in the text.

Figure 2. Event Study from Nov 2019 to Aug 2020.

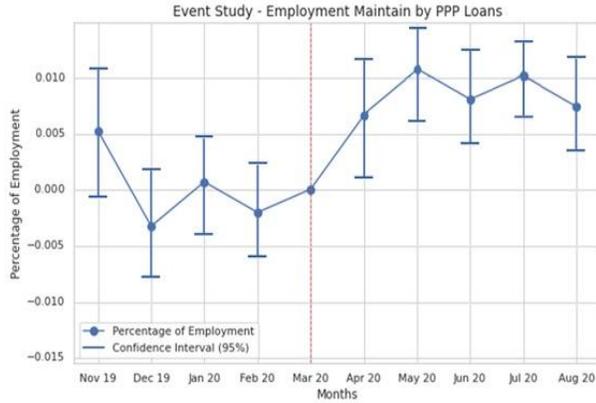
A Employment

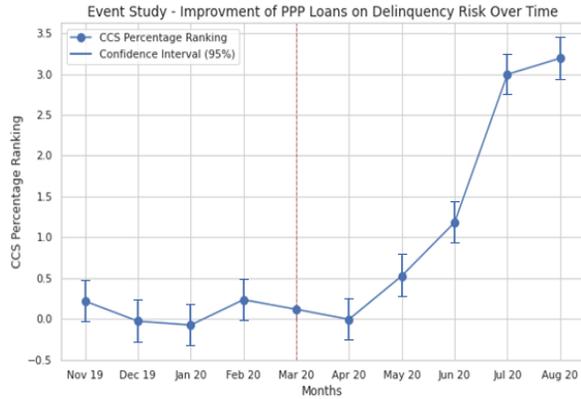
B. Commericial Credit Ranking

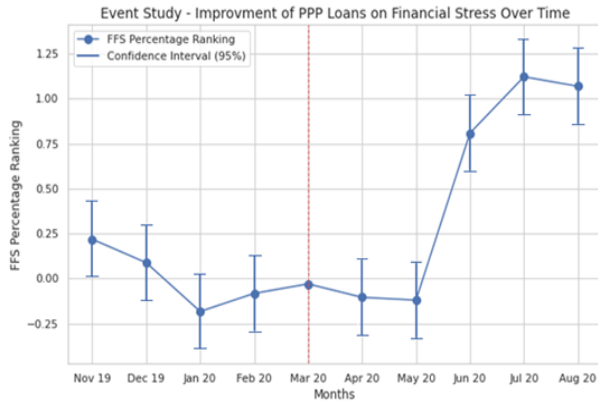
C. Financial Stress Ranking

*Notes:* These figures illustrate the DID event studies model (3), using panel data of size 3,681,052 from November 2019 to August 2020. March 2020 is treated as the category group. Panels A, B, and C examine the dynamic impact of PPP loans on log employment, commercial credit ranking, and financial stress ranking, respectively. The confidence bands are important to determine whether there were pre-existing trends in the outcome variables.
*Source:* Author's calculations based on Dun & Bradstreet datasets described in the text.

# Appendix A. Institutional Details of the PPP

## A.1. Important Events and Policy Adjustment of the PPP

The Paycheck Protection Program (PPP) was created under the CARES Act in March 2020 as a large-scale, forgivable loan program to support payroll and essential operating expenses for small and medium-sized businesses during COVID-19-related shutdowns (Hubbard and Strain, 2020; Autor et al., 2021a). The program operated in two broad windows. Window One in 2020 comprised two funding rounds. The First Round ran from April 3 to April 16, initially authorizing $342 billion across about 1.6 million loans. After these funds were quickly exhausted, the Second Round ran from April 27 to August 8 with an additional $180 billion, ultimately bringing the total number of first-draw loans in 2020 to about 5.1 million. Window Two in 2021 provided a Third Round of funding, including both first-draw loans for firms that had not previously participated and "second-draw" loans for firms that had already received PPP support. In this paper, I focus on loans disbursed in Window One in 2020.

First-draw PPP loans were generally available to businesses with 500 or fewer employees, but firms with more employees could qualify if they met SBA industry-specific size standards or were otherwise classified as "small business concerns" under existing statutes. Eligibility was not limited to traditional small businesses: tax-exempt nonprofit organizations, veterans' organizations, tribal businesses, housing cooperatives, and certain nonprofit news organizations could also apply. Borrowers could receive up to 2.5 times their average monthly payroll costs (3.5 times for accommodation and food services), subject to a cap of $10 million in 2020 and $2 million for second draws in 2021, at a fixed 1% interest rate and a five-year maturity. Loans were fully forgivable if borrowers maintained payroll and used at least 60% of the proceeds for eligible wage costs during the covered period. In practice, the loan size distribution was highly skewed: roughly two-thirds of recipients obtained loans of < $50,000.

The program's short-run objective (April–June 2020) was to prevent mass layoffs and business failures by subsidizing payroll, preserving employer–employee matches, and buffering extreme liquidity stress (Hubbard and Strain, 2020; Bartik et al., 2020). As lockdowns eased and additional funding became available, the program's role gradually shifted toward supporting ongoing operations and demand (Barrero et al., 2020; Miroudot, 2020). Subsequent rounds sought to correct perceived problems in initial implementation, including the "first-come, first-served" allocation mechanism and concerns about unequal access, by steering more resources toward smaller and more severely affected firms (Granja et al., 2020; Federal Reserve, 2020).

## A.2. Compliance of 500 Cutline Rule

The Paycheck Protection Program (PPP) has certain exceptions that allow businesses with more than 500 employees to be eligible for loans. According to the Small Business Administration (SBA), a business can be eligible for the PPP even if it has more than 500 employees if it satisfies the existing statutory and regulatory definition of a small business concern. Here are the key points:

1. Industry Size Standards: A business can qualify if it meets the SBA employee-based or revenue-based size standard corresponding to its primary industry. The size standards vary by industry and are based on the number of employees or annual receipts. You can find the industry size standards on the [SBA website](#).
2. Alternative Size Standard: A business can qualify for the PPP as a small business concern if it meets both tests in SBA's alternative size standard as of March 27, 2020: Maximum tangible net worth of the business is not more than $15 million.
The average net income after Federal income taxes (excluding any carry-over losses) of the business for the two full fiscal years before the date of the application is not more than $5 million.
3. Affiliation Rules: Borrowers must apply the affiliation rules set forth in SBA's Interim Final Rule on Affiliation. A borrower must certify that it is eligible to receive a PPP loan, and that certification means that the borrower is a small business concern as defined in section 3 of the Small Business Act, meets the applicable SBA employee-based or revenue-based size standard, or meets the tests in SBA's alternative size standard after applying the affiliation rules if applicable.
4. Non-Profit Organizations and Tribal Business Concerns: PPP loans are also available for qualifying tax-exempt nonprofit organizations, tax-exempt veterans' organizations, and Tribal business concerns that have 500 or fewer employees whose principal place of residence is in the United States, or meet the SBA employee-based size standards for the industry in which they operate. Source: [Paycheck Protection Program Loans Frequently Asked Questions](#)

# Appendix B – Dun & Bradstreet Data

## B.1. Failure Score (Formerly Financial Stress Score, FFS)

Dun & Bradstreet gathers data on businesses from many sources. Public records, financial accounts, and previous payment histories are a few examples. The Failure Score compares the demographics of a company to those of other businesses in the same sector. Among the information components utilized to determine the score are:

*Corporate Linkage* - When determining the Score for a specific organization, the size, strength, and risk of a group of enterprises are considered.
*Principals* - The experience and performance of affiliated enterprises of the principal.
*Financial* - Ratios and patterns gleaned from accounts monetary.
*Liquidity*, solvency, profitability, debt, late filing, and negative notes are among the factors evaluated.
*Company Type*: In general, corporations are thought to be less dangerous than other business types. These companies often have access to additional funding if needed.
*Age of the Company*: A company's stability can be gauged by how long it has been in operation. The danger decreases with the length of time a company has been in business.
*Open lawsuits, liens, and judgments* are indicated by their presence and quantity.
*Trade Data* - The Dun & Bradstreet Data Cloud model includes the proportion and monetary value of positive payment experiences. The danger is reduced by a higher percentage of positive trading experiences. Additionally, the likelihood of failure decreases as the overall number.

The Failure Score is a widely used tool for predicting a company's likelihood of failing. To offer additional predictive insight, it makes use of the entire range of Dun & Bradstreet data, including financials, comparative financial ratios, payment histories, public filings, demographic data, and more.

Some business associates, including suppliers, bankers, landlords, and consumers, could be curious in a company's Failure Score. To assist manage risk, businesses and financial institutions frequently use Dun & Bradstreet's credit scores. If your company has a low Failure Score, getting loans may be more difficult or have less favorable terms. You can also pass up profitable opportunities because you are worried about your company's capacity to handle its financial commitments.

## B.2. Delinquency Score (Formerly Commercial Credit Score)

To assess the likelihood that a company would become seriously delinquent on its accounts, Dun & Bradstreet performs predictive modeling analysis. These models are based on information that Dun & Bradstreet routinely gathers about companies, such as Trade References, public filings (lawsuits, liens, judgments, and bankruptcies), and financial statements. To predict the likelihood of future delinquency for 30 million US businesses, the Delinquency Predictor Score analyzes a company's past trade payment obligations (including granular payment data that captures month-to-month trends) that were submitted to Dun & Bradstreet, firmographic data, public information, and financial data. The Delinquency Predictor Score can even assist in providing a more accurate risk assessment for companies with little to no commercial transactions due to the wide range of data inputs.

Every time Dun & Bradstreet gathers additional pertinent information about a company or whenever information changes, the Delinquency Score is updated. For instance, a company's risk normally reduces with time, and the score may adjust annually to reflect this.

Even though other businesses also offer delinquency scores, only the D&B Delinquency Score has access to the depth of data from the Dun & Bradstreet Data Cloud that offers actionable predictive information and superior analytical methodology that makes it a more futuristic evaluation.

For companies that extend trade credit, the Delinquency Predictor Score can help assess potential risk without the need to conduct extensive research of their own. It provides guidance for setting credit limits, payment terms, and upfront deposit amounts.

Table B.1. Scale of Dun & Bradstreet Failure Score and Delinquency Score.

| Panel A D&B Failure Score (F Score) | | | |
|---|---|---|---|
| Failure Percentile | Failure Class | Incidence of Failure | Class Distribution |
| [95,100] | 1 | 0.03% | 6% |
| [69,94] | 2 | 0.09% | 26% |
| [34,-68] | 3 | 0.24% | 35% |
| [2,33] | 4 | 0.84% | 32% |
| 1 | 5 | 4.70% | 1% |
| Panel A D&B Delinquency Score (C Score) | | | |
| Failure Percentile | Failure Class | Incidence of Failure | Class Distribution |
| [95,100] | 1 | 1.00% | 6% |
| [69,94] | 2 | 2.50% | 26% |
| [34,-68] | 3 | 5.80% | 35% |
| [2,33] | 4 | 9.40% | 32% |
| 1 | 5 | 53.10% | 1% |

*Notes:* This set of tables show the scoring system of Failure Score and Delinquency Score by Dun & Bradstreet. The incidence of failure is predicative relative likelihood of business failure of delinquency in the next 12 months.
*Source:* Dun & Bradstreet

### B.3. DUNS Number

Stepping back a little to explain the role of DUNS Number, it is a unique nine-digit identifier for a business created by credit Dun & Bradstreet. The Data Universal Numbering System - or DUNS number - was created and copyrighted in 1962 by Dun & Bradstreet. Dun & Bradstreet is one of the three major business credit bureaus (Dun & Bradstreet, Equifax and Experian are also Certified Vendors with the Small Business Financial Exchange (SBFE)). DUNS numbers have become the standard numbering system to identify business entities across the globe.

A DUNS Number and company credit scores show similar information about a firm, just as the social security number does, such as its bank account, tax returns, credit score, homes, etc. The Delinquency Score and Failure Score, which make up Dun & Bradstreet's proprietary business credit score system, are linked to a company's DUNS number. Differently, a DUNS Number is used for business credit reporting purposes, whereas the federal tax ID number of a business (also called employer identification number, or EIN) is issued by the IRS and used for tax identification purposes. This data is used to assess the likelihood that the business will repay loans or other commercial debts in a timely manner.

The main benefit of having a DUNS Number is that it is essential to establishing and enhancing a firm's creditworthiness. As a result, requesting a DUNS Number is one of the initial steps in building a strong credit file for a company so that it can access more funding options and collaborate with more companies in the future. Applying for a DUNS number is a simple process, but it does require a business owner to first establish the business as a legal entity (LLC, C-Corp, S-Corp, etc) with the residency state. Although it is not mandatory to have a DUNS Number in all cases, businesses that have contracts with a government, seeking financing or building business credit history as necessity for business growth, or have networked businesses that involve import and export, are required to have DUNS number in most cases. For example, the Connecticut Office of the Arts began requiring a DUNS Number on all grant applications in 2017.[1]

### B.4. PAYDEX Score

---

[1] Grant application FAQs, Connecticut Office of the Arts. https://portal.ct.gov/DECD/Content/Arts-and-Culture/Funding_Opportunities/What-is-a-Duns-Number

The PAYDEX Score is a proprietary credit score developed by Dun & Bradstreet (D&B) that is used to assess the payment behavior of a business. It ranges from 1 to 100, with a higher score indicating better payment performance. The PAYDEX Score specifically evaluates how promptly a business pays its suppliers and creditors, based on the payment data in D&B's database. A score of 80, for instance, indicates that, on average, the company pays its bills on time, while a score of 100 suggests that it pays its bills 30 days ahead of terms. Conversely, a lower score indicates that the company tends to pay its bills late. Businesses and suppliers often use the PAYDEX Score to make decisions about extending trade credit or terms to other businesses. It is similar in concept to a personal credit score but is specifically for business transactions.

Table B.2. NAICS industry code.

| SIC2 | Description | NAICS-2 | Description |
|---|---|---|---|
| 01 | Agriculture | 11 | Agriculture, Forestry, Fishing and Hunting |
| 02 | Mining and | 21 | Mining, Quarrying, and Oil Gas Extract |
| 03 | Construction | 22 | Utilities |
| 04 | Manufacturing | 23 | Construction |
| 05 | Transportation | 31-33 | Manufacturing |
| 06 | Wholesale | 42 | Wholesale Trade |
| 07 | Retail | 44-45 | Retail Trade |
| 08 | Finance | 48-49 | Transportation and Warehousing |
| 09 | Services | 51 | Information |
| 10 | Public Administration | 52 | Finance and Insurance |
| | | 53 | Real Estate and Rental and Leasing |
| | | 54 | Professional, Scientific, and Technical Services |
| | | 55 | Management of Companies and Enterprises |
| | | 56 | Administrative and Support and Waste Management and Remediation Services |
| | | 61 | Educational Services |
| | | 62 | Health Care and Social Assistance |
| | | 71 | Arts, Entertainment, and Recreation |
| | | 72 | Accommodation and Food Services |
| | | 81 | Repair, Maintenance, Laundry Services, and Private Households |
| | | 92 | Public Administration |
| | | 99 | Unclassified Establishments |

*Notes:* This shows the 2-digit SIC Code and 2-digit NAICS Code of industry as controls for DID models. The 2-digit SIC Code is used for stratification towards heterogeneity study. The 2-digit NAICS Code is used for industry-by-month fixed effects.
*Source:* Dun & Bradstreet Data and BEA.

Table B.3. Correlations between PSM Features

|  | Ownership | Public | Minority | 7a History | log(sales) | $PPP_i$ |
|---|---|---|---|---|---|---|
| Ownership | 1 | 0.102 | 0.01 | 0.003 | -0.277 | 0.05 |
| Public | 0.102 | 1 | -0.009 | -0.007 | -0.237 | -0.031 |
| Minority | 0.01 | -0.009 | 1 | 0.022 | 0.028 | 0.059 |
| 7a History | 0.003 | -0.007 | 0.022 | 1 | 0.035 | 0.114 |
| log(sales) | -0.277 | -0.237 | 0.028 | 0.035 | 1 | 0.18 |
| $PPP_i$ | 0.05 | -0.031 | 0.059 | 0.114 | 0.18 | 1 |

*Notes:* This table shows the pairwise correlation between the key dependent variables and control variable, which helps consider the multicollinearity issue towards the selection of independent variables for constructing PSM model.
*Source:* Author's calculations based on Dun & Bradstreet datasets described in the text.

# Appendix C. Heterogenous Impact by 7(a) Experience and Industry

Table C.1. Stratified Impacts of the First Draw on 7(a) and Non-7(a) Borrowers

|  | Y = log(employment) | | Y = Financial Stress Ranking% | | Y = Commercial Credit% | |
|---|---|---|---|---|---|---|
| Mean of Y | 1.705 | 1.489 | 44.505 | 39.477 | 59.833 | 46.465 |
|  | (0.834) | (0.824) | (26.387) | (24.945) | (32.106) | (30.020) |
|  | (1) | (2) | (3) | (4) | (5) | (6) |
|  | 7(a) | Non-7(a) | 7(a) | Non-7(a) | 7(a) | Non-7(a) |
| *Panel A. Loan impact* | | | | | | |
| $PPP_i * Post$ | 0.0178*** | 0.0196*** | 0.900*** | 1.1046*** | 1.3113*** | 3.2282*** |
|  | (0.00) | (0.00) | (0.05) | (0.04) | (0.08) | (0.05) |
| Observations | 205,117 | 30,165,073 | 205,117 | 30,165,073 | 205,117 | 30,165,073 |
| $R^2$ | 0.004 | 0.196 | 0.011 | 0.203 | 0.01 | 0.192 |
|  | (1) | (2) | (3) | (4) | (5) | (6) |
|  | 7(a) | Non-7(a) | 7(a) | Non-7(a) | 7(a) | Non-7(a) |
| *Panel B. Duration impact* | | | | | | |
| Duration | 0.0038*** | 0.0038*** | 0.1872*** | 0.1692*** | 0.2684*** | 0.5699*** |
|  | (0.00) | (0.00) | (0.01) | (0.00) | (0.02) | (0.00) |
| Observations | 0 | 30,165,073 | 205,117 | 30,165,073 | 205,117 | 30,165,073 |
| $R^2$ | 0.004 | 0.013 | 0.011 | 0.002 | 0.01 | 0.002 |

*Notes:* This table reports heterogeneous effects of the first PPP draw by firms' prior SBA 7(a) borrowing status. Columns (1)–(2) use log(employment) as the dependent variable, columns (3)–(4) use the percentile rank of the D&B Financial Stress Score (higher values indicate lower predicted failure risk), and columns (5)–(6) use the percentile rank of the D&B Commercial Credit Score (higher values indicate better commercial credit quality. "7(a)" firms are establishments with a documented SBA 7(a) lending history prior to the PPP; "Non-7(a)" firms have no such history. Panel A reports DID estimates for an indicator that equals one for PPP recipients in the post-treatment period (September 2020) and zero otherwise (March 2020). Panel B reports the associated coefficients on loan duration (measured in years). All specifications include industry, state, and firm fixed effects. Robust standard errors are clustered at the firm level. ***, **, and * denote significance at the 1%, 5%, and 10% levels, respectively.
*Source:* Author's calculations based on Dun & Bradstreet datasets described in the text.

Table C.2(a) Effects of the First Draw on Employment Stratified by NAICS2

| | Dependent Variable: Y = log(employment) | | | | | | | | |
|---|---|---|---|---|---|---|---|---|---|
| Mean of Y | 1.039 | 1.225 | 1.874 | 1.486 | 1.214 | 1.753 | 1.878 | 1.230 | 1.946 |
| | (0.557) | (0.766) | (1.276) | (0.881) | (0.708) | (1.007) | (0.917) | (0.620) | (0.467) |
| | (1) | (2) | (3) | (4) | (5) | (6) | (7) | (8) | (9) |
| | Agric | Mng/Cnstr | Manufact | Retail/Trsp | Fin/RE/Tech | Edu/Hlth | Food/Ent | Repair/HH | Public Adm |
| *Panel A. Loan impact* | | | | | | | | | |
| $PPP_i * Post$ | 0.0101*** | 0.021*** | 0.041*** | 0.027*** | 0.017*** | 0.027*** | 0.034*** | 0.017*** | (0.001) |
| | (0.001) | (0.000) | (0.001) | (0.001) | (0.001) | (0.001) | (0.000) | (0.001) | (0.001) |
| Observations | 529,735 | 2,283,346 | 882,257 | 4,502,777 | 9,377,352 | 2,806,939 | 1,778,571 | 2,841,530 | 5,360,403 |
| $R^2$ | 0.014 | 0.013 | 0.015 | 0.023 | 0.013 | 0.012 | 0.016 | 0.014 | 0.011 |
| | (1) | (2) | (3) | (4) | (5) | (6) | (7) | (8) | (9) |
| | Agric | Mng/Cnstr | Manufact | Retail/Trsp | Fin/RE/Tech | Edu/Hlth | Food/Ent | Repair/HH | Public Adm |
| *Panel B. Duration impact* | | | | | | | | | |
| Duration | 0.002*** | 0.005*** | 0.009*** | 0.006*** | 0.004*** | 0.006*** | 0.007*** | 0.004*** | -0.0002*** |
| | (0.000) | (0.000) | (0.000) | (0.000) | (0.000) | (0.000) | (0.000) | (0.000) | (0.000) |
| Observations | 529,735 | 2,283,346 | 882,257 | 4,502,777 | 9,377,352 | 2,806,939 | 1,778,571 | 2,841,530 | 5,360,403 |
| $R^2$ | 0.014 | 0.013 | 0.015 | 0.019 | 0.017 | 0.024 | 0.016 | 0.024 | 0.001 |

Table C.2(b) Effects of the First Draw on Financial Stress Stratified by NAICS2

| | Dependent Variable: Y = Financial Stress Percentile Ranking | | | | | | | | |
|---|---|---|---|---|---|---|---|---|---|
| Mean of Y | 52.234 | 39.547 | 41.914 | 38.147 | 43.719 | 44.226 | 38.600 | 44.891 | 26.616 |
| | (28.292) | (25.859) | (26.806) | (25.740) | (24.516) | (26.085) | (25.182) | (25.924) | (16.507) |
| | (1) | (2) | (3) | (4) | (5) | (6) | (7) | (8) | (9) |
| | Agric | Mng/Cnstr | Manufact | Retail/Trsp | Fin/RE/Tech | Edu/Hlth | Food/Ent | Repair/HH | Public Adm |
| *Panel A. Loan impact* | | | | | | | | | |
| $PPP_i * Post$ | 0.686*** | 0.528*** | 0.877*** | 1.587*** | 1.289*** | 0.909*** | 1.778*** | 1.114*** | 2.377*** |
| | (0.033) | (0.015) | (0.024) | (0.037) | (0.036) | (0.036) | (0.019) | (0.038) | (0.056) |
| Observations | 529,735 | 2,283,346 | 882,257 | 4,502,777 | 9,377,352 | 2,806,939 | 1,778,571 | 2,841,530 | 5,360,403 |
| $R^2$ | 0.003 | 0.001 | 0.004 | 0.012 | 0.009 | 0.006 | 0.011 | 0.007 | 0.006 |
| | (1) | (2) | (3) | (4) | (5) | (6) | (7) | (8) | (9) |
| | Agric | Mng/Cnstr | Manufact | Retail/Trsp | Fin/RE/Tech | Edu/Hlth | Food/Ent | Repair/HH | Public Adm |
| *Panel B. Duration impact* | | | | | | | | | |
| Duration | 0.149*** | 0.109*** | 0.180*** | 0.322*** | 0.237*** | 0.153*** | 0.379*** | 0.176*** | 0.483*** |
| | (0.007) | (0.003) | (0.005) | (0.003) | (0.002) | (0.003) | (0.004) | (0.003) | (0.003) |
| Observations | 529,735 | 2,283,346 | 882,257 | 4,502,777 | 9,377,352 | 2,806,939 | 1,778,571 | 2,841,530 | 5,360,403 |
| $R^2$ | 0.003 | 0 | 0.004 | 0.008 | 0.003 | 0.001 | 0.011 | 0.001 | 0.077 |

Table C.2(c) Effects of the First Draw on Commercial Credit Stratified by NAICS2

| | Dependent Variable: Y = Commercial Credit Percentile Ranking | | | | | | | | |
|---|---|---|---|---|---|---|---|---|---|
| Mean of Y | 71.913 | 45.521 | 55.929 | 48.098 | 45.118 | 58.718 | 46.173 | 56.389 | 32.722 |
| | (23.754) | (32.048) | (31.633) | (32.134) | (29.688) | (28.203) | (31.221) | (30.231) | (20.515) |
| | (1) | (2) | (3) | (4) | (5) | (6) | (7) | (8) | (9) |
| | Agric | Mng/Cnstr | Manufact | Retail/Trsp | Fin/RE/Tech | Edu/Hlth | Food/Ent | Repair/HH | Public Adm |
| **Panel A. Loan impact** | | | | | | | | | |
| $PPP_i$ * Post | 1.113*** | 2.086*** | 1.654*** | 3.503*** | 3.658*** | 1.219*** | 3.149*** | 2.480*** | 6.477*** |
| | (0.037) | (0.023) | (0.040) | (0.059) | (0.053) | (0.056) | (0.031) | (0.056) | (0.071) |
| Observations | 529,735 | 2,283,346 | 882,257 | 4,502,777 | 9,377,352 | 2,806,939 | 1,778,571 | 2,841,530 | 5,360,403 |
| $R^2$ | 0.001 | 0.002 | 0.005 | 0.009 | 0.008 | 0.004 | 0.005 | 0.004 | 0.002 |
| | (1) | (2) | (3) | (4) | (5) | (6) | (7) | (8) | (9) |
| | Agric | Mng/Cnstr | Manufact | Retail/Trsp | Fin/RE/Tech | Edu/Hlth | Food/Ent | Repair/HH | Public Adm |
| **Panel B. Duration impact** | | | | | | | | | |
| Duration | 0.237*** | 0.430*** | 0.338*** | 0.704*** | 0.712*** | 0.186*** | 3.149*** | 0.416*** | 1.455*** |
| | (0.008) | (0.005) | (0.009) | (0.004) | (0.003) | (0.005) | (0.031) | (0.000) | (0.004) |
| Observations | 529,735 | 2,283,346 | 882,257 | 4,502,777 | 9,377,352 | 2,806,939 | 1,778,571 | 2,841,530 | 5,360,403 |
| $R^2$ | 0.001 | 0.002 | 0.004 | 0.007 | 0.004 | 0.001 | 0.005 | 0.001 | 0.03 |

*Notes:* This table reports heterogeneous effects of the first PPP draw on establishment employment, financial stress, and delinquency risk, by NAICS 2-digit industry groups. All the three tables show the results of model (1) and (2), specifically, 2(a) shows the effects of the first draw on log employment, while 2(b) and 2(c) are the results for financial stress percentile ranking and commercial credit percentile ranking, respectively. Within each table, Panel A presents DID estimates for the PPP treatment indicator. Panel B reports the corresponding coefficients on loan duration (measured in years). The sample contains two periods: March 2020 (Pre = 0) and September 2020 (Post = 1). Industry headings correspond to the following NAICS groupings: Agriculture (11); Mng, Cnstr (21–23); Manufact (31–33); Retail, Trsp (44–45, 48–49); Fin, RE, Tech (finance, insurance, real estate, and information/technology services; 52–53, 51); Edu, Hlth (61–62); Food, Ent (accommodation, food services, and arts, entertainment, and recreation; 72, 71); Repair, HH (other services and household services; 81); and Public Adm (92). All specifications include industry, state, and firm fixed effects. Robust standard errors are clustered at the firm level. ***, **, and * denote significance at the 1%, 5%, and 10% levels, respectively.
*Source:* Author's calculations based on Dun & Bradstreet datasets described in the text.